\documentclass[twocolumn,showpacs,preprintnumbers,amsmath,amssymb,prb]{revtex4}
\usepackage{graphicx}
\usepackage{dcolumn}
\usepackage{bm}
\usepackage{float}

\begin{document}

\title{Non-trivial spin-texture of the coaxial Dirac cones on the surface of topological crystalline insulator SnTe}
\author{Yung Jui Wang$^{1,2}$, Wei-Feng Tsai$^3$, Hsin Lin$^1$, Su-Yang Xu$^4$, M. Neupane$^4$, M.Z. Hasan$^4$, and A. Bansil$^1$}
\affiliation{$^1$Department of Physics, Northeastern University, Boston, Massachusetts 02115, USA \\
$^2$Advanced Light Source, Lawrence Berkeley National Laboratory, Berkeley, California 94305, USA \\
$^3$Department of Physics, National Sun Yat-sen University, Kaohsiung 80424, Taiwan
$^4$Joseph Henry Laboratory, Department of Physics, Princeton University, Princeton, New Jersey 08544, USA \\
}

\date{\today}

\begin{abstract}
We present first principles calculations of the nontrivial surface states and their spin-textures in the topological crystalline insulator SnTe.
The surface state dispersion on the [001] surface exhibits four Dirac-cones centered along the intersection of the mirror plane and the surface 
plane. We propose a simple model of two interacting coaxial Dirac cones to describe both the surface state dispersion and the associated 
spin-texture. While the out-of-the-plane spin polarization is zero due to the crystalline and time-reversal symmetries, 
the in-plane spin texture shows helicity with some distortion due to the interaction of the two coaxial Dirac cones, 
indicating a nontrivial mirror Chern number of -2, distinct from the value of -1 in $Z_{2}$ topological insulator such as 
Bi/Sb alloys or Bi$_2$Se$_3$. The surface state dispersion and its spin-texture would provide an experimentally 
accessible way to determine the nontrivial mirror Chern number.

\end{abstract}

\pacs{73.20.r, 73.43.Cd}
\maketitle

\section{Introduction}
Since the discovery of time-reversal symmetry protected topological quantum
states in two-dimensional (2D) Hg(Cd)Te-based quantum well structures, and subsequently 
that of the Z$_{2}$ three-dimensional (3D) topological insulators \cite{reviewHasan,reviewZhang,reviewMoore,scHgTe,FuBi,HsiehSb,HsiehBi}, 
an enormous effort has been dedicated to finding other novel materials, which could support non-trivial topological states. 
One particularly fruitful recent direction has been to explore quantum states in condensed matter systems, which are protected 
by the symmetries of the lattice, leading to the so-called topological crystalline insulators (TCIs).\cite{FuTci} A practical 
realization of a TCI has been the prediction of SnTe with an ideal rocksalt structure in which the mirror symmetry of the lattice 
ensures the existence of robust metallic edge states.\cite{Natliang}. This theoretical prediction was verified quickly via angle 
resolve photoemission experiments on Pb$_{1-x}$Sn$_{x}$Te
\cite{ARSuYang,ARTanaka} and Pb$_{1-x}$Sn$_{x}$Se\cite{ARDziawa}. 
Recall that a Z$_{2}$ topological insulator such as Bi$_{2}$Se$_{3}$/Te$_{3}$, which is protected by time-reversal 
symmetry, contains a single Dirac cone at the center of the [111] surface plane, exhibiting 
a linear dispersion and chiral spin texture predicted theoretically as well as observed experimentally \cite{HsiehBi}. In contrast 
to the 3D Z$_{2}$ topological insulators, however, a TCI protected by mirror symmetry contains an even number of Dirac cones on 
crystal surfaces symmetric about the [110] mirror planes, and its topologically non-trivial state is characterized by a nonzero 
mirror Chern number\cite{PRBTeo,Natliang}. The interest in understanding the bizarre surface states and their spin-textures in 
SnTe, the pristine phase of an archetype TCI, is thus obvious, both in its own right and as a way of gaining insight into the 
properties of related substitutional compounds.

\begin{figure}
\includegraphics[width=8cm]{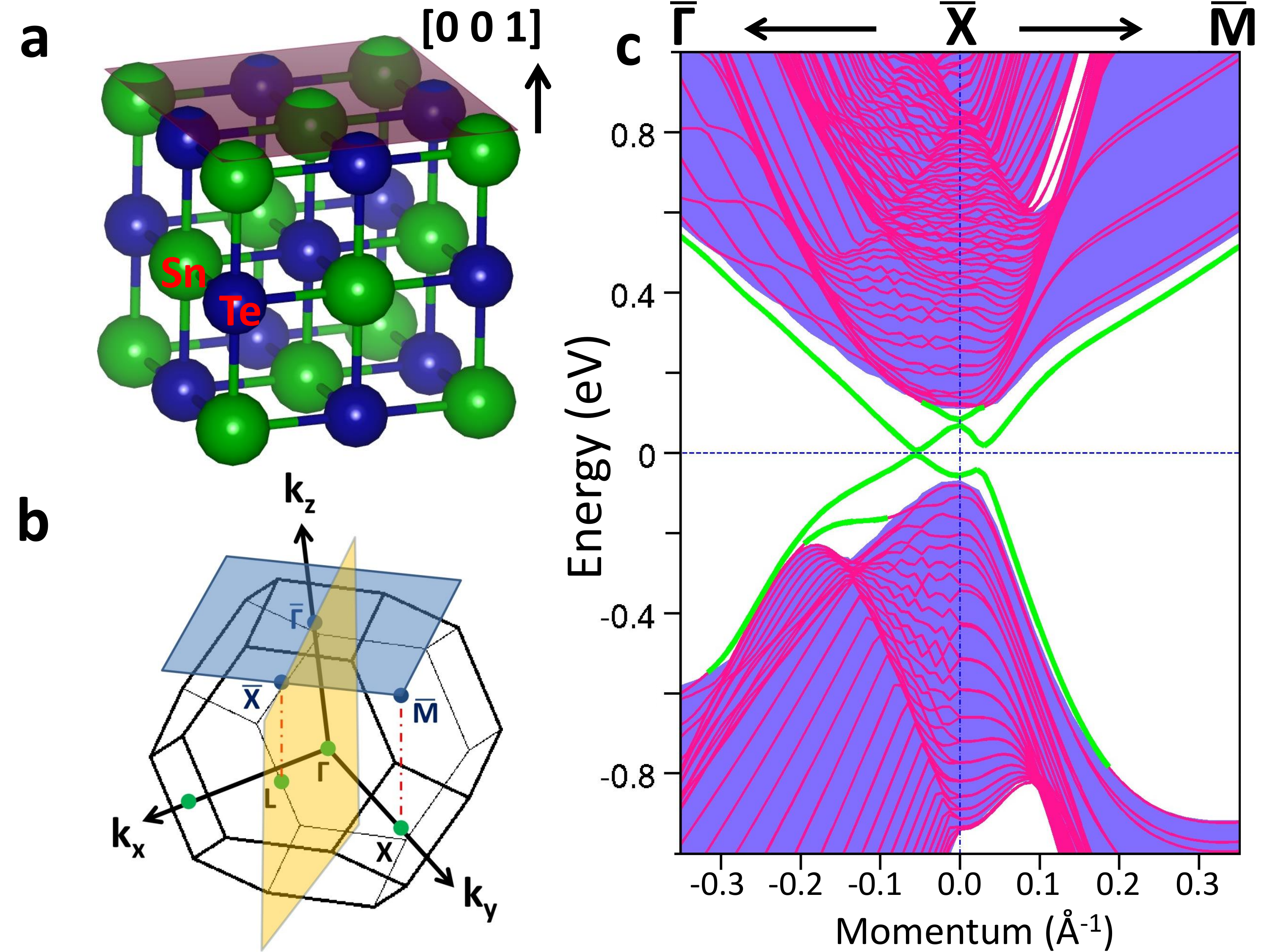}
\caption{(Color online)
{\bf Crystal structure and surface bands of SnTe.}
{\bf a.} Rocksalt crystal structure of SnTe. A [001] surface plane is shown.
{\bf b.} FCC Brillouin zone (BZ) of SnTe and the [001] surface BZ. 
The [1 -1 0] mirror plane crosses the 2D surface BZ along  $\bar{\Gamma}$\={X}.
{\bf c.} Electronic structure of SnTe [001] surface around \={X}. The 
surface bands are indicated by thick green lines and bulk bands 
by the shaded purple area. 
A Dirac point of surface state appears 
along the $\bar{\Gamma}$\={X} direction, while the surface band is 
gapped along $\bar{X}$\={M}. 
}
\label{f1}
\end{figure}
%

\begin{figure*}
\includegraphics[width=14cm]{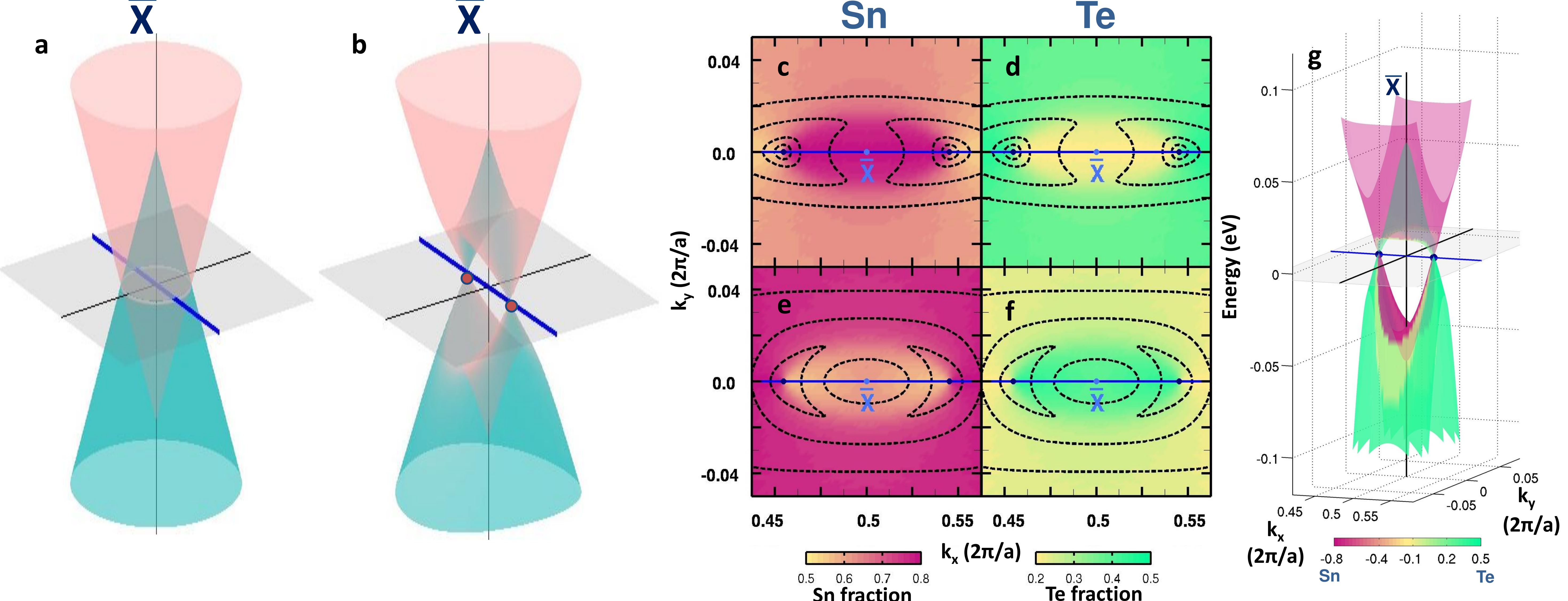}
\caption{(Color online)
{\bf Two interacting coaxial Dirac cones.}
 Schematic diagram for a non-interacting (interacting) coaxial Dirac cone centered at the \={X} point
is shown in {\bf a} ({\bf b}).
As the interaction between the two Dirac cones is turned on, 
gaps open up except at the two new Dirac points on the two sides of the \={X} point. The new gapless Dirac cones are 
protected by the presence of the mirror plane which intersects the surface BZ along $\bar{\Gamma}$\={X} (bold blue line). 
Maps of the fraction of the partial {\bf charges} on Sn and Te atoms
for the {\bf valence} surface bands are shown in {\bf c} and {\bf d}, respectively, while those for the conduction 
surface bands are shown in {\bf e} and {\bf f}. Dashed lines are the constant energy contours.
{\bf g.} Surface band dispersion with colors indicating the fraction of partial charges on the Sn and Te atoms.}
\label{f2}
\end{figure*}

In this work, we report first-principles calculations to investigate the [001] surface states of SnTe in rocksalt structure. 
By including the spin-orbit coupling, two gapless Dirac cones centered 
along $\bar{\Gamma}$\={X} (the other two lie along $\bar{\Gamma}$\={Y}), a mirror line of the crystal symmetry, are 
found in the surface spectrum, a result of the nontrivial band topology due to crystal symmetry. We examine the complicated 
surface band characters as well as the nontrivial spin textures, and propose a simple model which consists of two interacting 
coaxial Dirac cones centered at \={X}.
When the interaction between the two coaxial Dirac cones is turned on, 
the two surface bands avoid each other and a gap opens up except on 
the mirror line, forming the gapless Dirac cones centered along $\bar{\Gamma}$\={X}. While the out-of-the-plane spin 
polarization vanishs due to C$_{4v}$ symmetry as well as the superposition of the mirror and time reversal symmetries, 
the in-plane spin textures show helicity with some distortion due the interaction of the two Dirac cones. The overall 
spin texture reveals the nontrivial mirror Chern number with the value of -2, which is distinct from that of -1 
in $Z_{2}$ topological insulators such as Bi/Sb alloy or Bi$_2$Se$_3$.\cite{HsiehSb,PRBTeo,TakahashiPRL} The surface 
state dispersion and the associated spin-texture would provide an experimentally accessible way to determine the 
nontrivial mirror Chern number.

\section{Band structure calculations}
We first extract both the electronic band structure and the spin texture of the SnTe surface states from first-principles 
calculations, which was carried out within the framework of the density 
functional theory (DFT) using pseudo-potential projected augmented wave 
method\cite{paw} as implemented in the VASP package \cite{vasp}. The 
generalized gradient approximation (GGA)\cite{gga} was used to model
exchange-correlation effects. The spin orbital coupling (SOC) is 
included in the self-consistent cycles. The surfaces are modeled by 
periodically repeated slabs of 33-atomic-layer thickness with 
13 angstrom wide vacuum regions, using a 12x12x1 Monkhorst-Pack 
k-point mesh over the Brillouin zone (BZ) with 208 eV cutoff energy. 
The room temperature crystal structure of SnTe in an ideal sodium chloride
structure was used to construct the slab without any rhombohedral distortion. The experimental lattice constant of 
SnTe with the value of 6.327 \AA{} was 
used\cite{lasnte}. The self-consistent Bloch wavefunctions associated with 
the surface states were decomposed into cubic spherical harmonic orbitals 
and projected on to various atomic sites. For each atomic site in SnTe,
we obtained the charge density and the three components of the 
spin direction from the expectation 
values of the $x$, $y$ and $z$ Pauli matrices.

In order to clarify the relevant structural aspects, SnTe in an ideal sodium chloride FCC crystal structure is shown 
in Fig.~1a. Fig.~1b shows the FCC bulk Brillouin zone (BZ) as well as the [001] surface BZ. The high symmetry 
points $\Gamma$, L and X in the bulk BZ are projected on the $\bar{\Gamma}$, \={X} and \={M} in the surface 
BZ, respectively. The [1 -1 0] mirror plane is perpendicular to the [001] surface plane and projected onto the 
surface BZ along the $\bar{\Gamma}$\={X} direction. The projected bulk band structure along the high symmetry 
lines $\bar{\Gamma}$-\={X}-\={M} is shown in Fig.~1c by purple area. 
The surface bands are highlighted by thick green lines, isolated from other bulk bands. A gapless Dirac cone with 
the Dirac point sits along $\bar{\Gamma}$-\={X} 
on the two sides of the \={X} point at the Fermi level, while the surface states along \={X}-\={M} are gapped.

The complicated surface states of SnTe can be understood by a simpler picture, which consists of 
two Dirac cones. Fig.~2a shows two coaxial Dirac cones centered at the \={X} point. 
The \={X} and \={Y} points, i.e. (0.5, 0) and (0, 0.5), are equivalent and we concentrate on the 
one at (0.5,0) hereafter. These two Dirac cones carry different band characters and can be distinguished 
through an analysis of the wavefunctions of the Sn and Te atoms on the surface layers of the [001] slab. 
The nature of the differences in potentials of the Sn and Te atoms separates the two Dirac cones in energy vertically as shown in 
Fig.~2a. If there were no interactions between the two Dirac cone states, 
the two Dirac cones will intersect each other in the 2D surface BZ in an ellipse, 
which only intersects the mirror line along $\bar{\Gamma}$-\={X} (blue line)
with two points on the two sides of the \={X} point. In Fig.~2b, when the interaction between the 
two Dirac cones is turned on, a gap opens up along the elliptical overlap region of the 
two Dirac cones, except at the two points protected by the mirror symmetry.

In order to trace the band character of the SnTe surface
states, we decompose their charge distribution into Sn 
and Te partial contributions presented in Figs.~2c-f. 
For the surface valence states (Fig.~2c), we find that the Sn fraction is 
dominant within an elliptical region between the two Dirac points and centered at the \={X} point, while 
in Fig.~2d the Te fraction is larger outside this region. But, for the conduction 
surface states (Figs.~2e and f), this trend between the partial contributions from Se and Te atoms is reversed. 
To enhance visual clarity, we patch the charge fraction map onto the energy dispersion in Fig.~2g.
These results of first-principles calculations qualitatively agree 
with the picture of two interacting Dirac cones shown in Fig.~2b 
and allow us to attribute the origin of the two coaxial Dirac cones to distinct atomic orbitals. 
One of the cones with opening to the high energy side is more Sn-like with $p_z$-orbital, while the other 
cone with opening to the low energy side is more Te-like with $p_x$-orbital.
The Sn-like cone has the \={X}-point Dirac point lower in energy than
that of the Te-like cone.

\begin{figure}
\includegraphics[width=8.5cm]{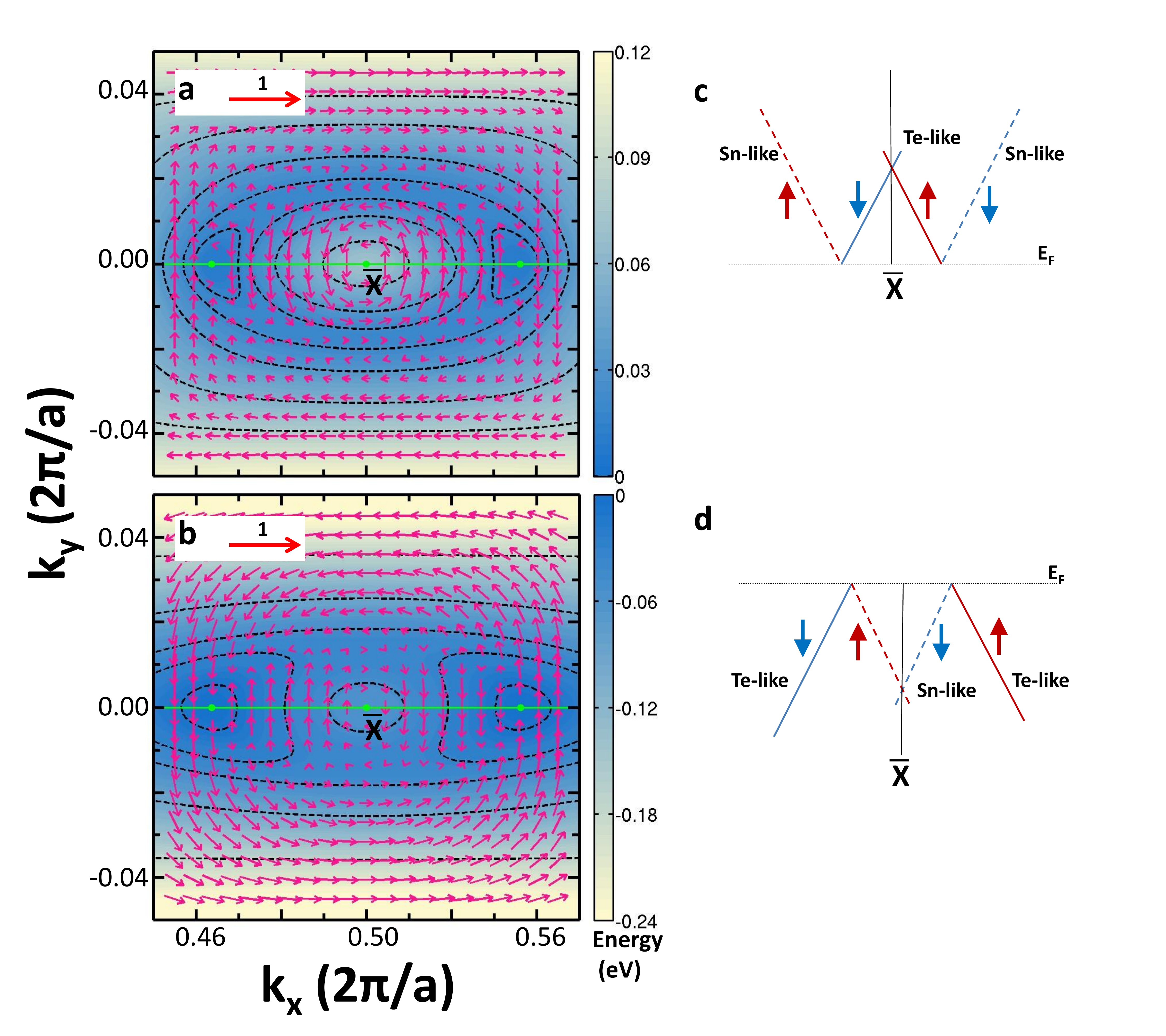}
\caption{(Color online)
{\bf Non-trivial spin-texture of SnTe surface states.}
Spin textures of the conduction and valence surface bands are shown in {\bf a} and {\bf b}, respectively. 
Energy is represented by the color scale, and constant energy contours are shown.
Green line indicates the $\bar{\Gamma}$\={X} direction and green dots denote the Dirac points.
Schematic band structures for the conduction and valence surface bands along $\bar{\Gamma}$\={X} direction 
are shown in {\bf c} and {\bf d}, respectively. Solid lines are Te-like Dirac cones and the dashed lines 
are Sn-like. Red (blue) arrows indicate the spin along positive (negative) y-direction.
}
\label{f3}
\end{figure}

In Fig.~3, we plot surface band energies using a color scale and the associated constant energy contours, together with the in-plane spin-texture.
The spin direction of each state was obtained here by calculating the expectation of the Pauli matrices from the first 6 atomic 
layers from the top surface of the SnTe slab. For the valence surface states shown in Fig.~3b, the highest energy state is 
located at the Dirac points lying along the $\bar{\Gamma}$\={X} mirror line. The constant energy contours close to the Dirac 
points are seen not to be of a perfect circular shape, which implies an anisotropic Fermi velocity 
around the Dirac points. Between two Dirac points, an energy valley is centered at the \={X} point. 
The energy contours exhibit a Lifshitz transition\cite{Natliang}. As we go to energies below the Fermi energy, 
the constant energy surface changes its topology. The two disconnected hole pockets next to the \={X} point at 
high energy become a large hole and a small electron pocket both centered at the \={X} point at low energy. 
In the two coaxial Dirac cones, the large hole pocket is associated with the Te-like lower Dirac cone and 
the small electron pocket is associated with the Sn-like upper Dirac cone. A similar change in the Fermi 
surface topology occurs in the conduction surface bands (Fig.~3a). At high energies above 
the interaction region, the large electron and small hole pocket centered at the \={X} point are associated
with the Sn-like upper Dirac cone and the Te-like lower Dirac cone, respectively. 

It is interesting to see how the spin-textures within our picture of two co-axial Dirac cones play out based on 
the spin-textures derived from our first-principles computations. Note that for the conduction bands shown 
in Fig.~3a, the spin texture shows counter-clockwise rotation around the \={X} point and clockwise rotation far away from the \={X} point.
As shown in the schematic diagram of Fig.~3c, states near the \={X} point are associated with the Te-like lower Dirac cone while those 
far away from the \={X} point belong to the Sn-like upper Dirac cone.
The two coaxial cones therefore should mimic similar chirality.
They should both have clockwise spin rotation in the upper cone and counter-clockwise in the lower cone. 
Along the $\bar{\Gamma}$\={X} direction, these two opposite spin states meet at the Dirac points. 
Along the \={X}\={M} direction, the spin polarization diminishes and switches direction around 
the Lifshitz transition. The spin texture of the valence surface states can be understood along 
similar lines. A clockwise rotation in the inner region around \={X} is associated with the Sn-like 
upper Dirac cone, while a counter-clockwise rotation in the outer region far away from the \={X} point 
is associated with the Te-like lower Dirac cone. We note that the spin chirality of the surface Dirac 
cone in a typical strong topological insulator like Bi$_{2}$Se$_{3}$\cite{HsiehBi} is the same as that 
of the Sn-like and Te-like Dirac cones we have found here on the surface of SnTe, and 
bears a nontrivial mirror Chern number further discussed below.

\section{Simplified $k\cdot p$ model}
In order to make our proposed two-coaxial-cone picture more concrete, 
we now discuss a 2D $\mathbf{k}\cdot\mathbf{p}$ model, which not only captures correctly 
the evolution in band dispersion under a Lifshitz transition, but also describes the spin 
texture reasonably\cite{foot1,ARFang}.
In this connection, it is natural to 
consider a minimal four-band model with Hamiltonian $H(k_x,k_y)$ around \={X} point on 
the [001] surface, which obeys the following three symmetries: mirror symmetry about 
the $xz$ plane ($M_{xz}$), mirror symmetry about the $yz$ plane ($M_{yz}$), and 
time-reversal symmetry ($\Theta=TK$, where $K$ for denotes complex conjugate). 
We then have  under these 
symmetry operations, $M_{xz}H(k_x,k_y)M^{-1}_{xz} = H(k_x,-k_y)$, $M_{yz}H(k_x,k_y)M^{-1}_{yz} = H(-k_x,k_y)$ 
and $TH(k_x,k_y)T^{-1} = H^*(-k_x,-k_y)$.

As shown in Fig.~2a, two distinct Dirac points occur at \={X}, associated with 
energies $E_+>E_0$ and $E_-<E_0$. [$E_0$ denotes the energy level at which the 
two cones intersect.] To account for the doublet state at each Dirac point, we 
choose the basis set to be the eigenvectors of $M_{yz}$ with 
eigenvalues $\pm i$: \{$|i;\text{Sn}\rangle$, $|-i;\text{Sn}\rangle$, $|-i;\text{Te}\rangle$, $|i;\text{Te}\rangle$\}, where 
the main atomic portion for each cone is indicated. In particular, when combined with the dominant orbitals 
for Sn ($p_z$) and Te ($p_x$) atoms mentioned earlier, this basis set also captures spin information, resulting 
in \{$|p_z,\rightarrow;\text{Sn}\rangle$, $|p_z,\leftarrow;\text{Sn}\rangle$, $|p_x,\rightarrow;\text{Te}\rangle$, $|p_x,\leftarrow;\text{Te}\rangle$\}. 
Note that the quantization axis for spin is now along $x$, with $M_{yz}|p_z,\rightarrow(\leftarrow)\rangle = \pm i |p_z,\rightarrow(\leftarrow)\rangle$ and $M_{yz}|p_x,\rightarrow(\leftarrow)\rangle = \mp i |p_x,\rightarrow(\leftarrow)\rangle$. Defining $4\times 4$ matrices,  $\Sigma_{\alpha\beta}=s_{\alpha}\otimes\sigma_{\beta}$, 
with Pauli matrices $\mathbf{s}$ and $\mathbf{\sigma}$ acting on spin and orbital spaces, respectively, the symmetry operator $M_{yz}$ then takes the form, $i\Sigma_{33}$ and the other two symmetry operators can be written as $M_{xz}=-i\Sigma_{20}$ and $T=-i\Sigma_{20}$. 
After examining the 16 $\Sigma$ matrices under all three symmetry operations, up to linear coupling in $k_x$, $k_y$, we obtain the following symmetry-allowed Hamiltonian,
\begin{eqnarray}
H(\vec{k})&=& m\Sigma_{03}+m^{\prime}\Sigma_{22} \nonumber \\
&+& k_x(v_{1x}\Sigma_{20}+v_{2x}\Sigma_{02}+v_{3x}\Sigma_{23}) \nonumber \\
&+& k_y(v_{1y}\Sigma_{30}+v_{2y}\Sigma_{11}+v_{3y}\Sigma_{33}).
\end{eqnarray}

In Fig.~2b, we plot the energy dispersions based on this effective Hamiltonian, which 
is clearly seen to capture the key features of first-principles calculations (See Fig.~2g). 
The presence of two new Dirac points is due to the double degeneracy given by 
different $M_{xz}$ eigenvalues and is thus protected by the mirror symmetry of the 
system about the $xz$ plane, in sharp contrast to the Dirac points at \={X}, 
which are mainly protected by the time-reversal symmetry. As to the spin texture, 
note that $\langle s_3\rangle$ and $\langle s_2\rangle$ now represent the in-plane 
spin $x$ and $y$ components, respectively. Furthermore, one can prove that for any 
eigenstate with momentum $\mathbf{k}$, $\langle \mathbf{k}|s_1|\mathbf{k}\rangle = 0$ (out-of-the-plane component), 
as required by the combined TRS and the two mirror 
symmetries: $\langle \mathbf{k}|s_1|\mathbf{k}\rangle = \langle \mathbf{k}|M_{xz}M_{yz}\Theta s_1\Theta^{-1}M_{yz}^{-1}M_{xz}^{-1}|\mathbf{k}\rangle = -\langle \mathbf{k}|s_1|\mathbf{k}\rangle$. 
It turns out that the resulting spin texture is qualitatively similar to that shown in Fig.~3. 
Finally, we note that if one applies a unitary transformation, $U=e^{i\frac{\pi}{4}s_2}\otimes\sigma_0$, to Eq. (1), 
the transformed $H(\vec{k})$ becomes
\begin{widetext}
\begin{equation}
\tilde{H}(\vec{k})=
\left(\begin{array}{cccc}
m & -iv_{x+}k_x-v_{y+}k_y & -iv_{2x}k_x+v_{2y}k_y & -m^\prime\\
iv_{x+}k_x-v_{y+}k_y & m & m^\prime & -iv_{2x}k_x-v_{2y}k_y \\
iv_{2x}k_x+v_{2y}k_y & m^\prime & -m & -iv_{x-}k_x-v_{y-}k_y\\
-m^\prime & iv_{2x}k_x-v_{2y}k_y & iv_{x-}k_x-v_{y-}k_y & -m
\end{array} \right)
\end{equation},
\end{widetext}
where $v_{x\pm}=v_{1x}\pm v_{3x}$ and $v_{y\pm}=v_{1y}\pm v_{3y}$. This is in fact a more familiar Hamiltonian, representing 
two interacting Dirac cones in the Rashba form, $\mathbf{k}\times\mathbf{s}$, with an identical chirality\cite{foot2,Fukpmodel}.
In Table.~I, we list the parameter sets used in Eq.~(2)
by fitting experimentally obtained ARPES band dispersions in
topologically nontrivial (Pb,Sn)Te\cite{ARSuYang}, SnTe\cite{ARTanaka,ARTanaka2}
and (Pb,Sn)Se\cite{ARDziawa,ARMadhabN}.

\begin{table}[h!b!p!]
\begin{center}
\caption{Parameter sets for the 2D $\mathbf{k}\cdot\mathbf{p}$ model
Hamiltonian of Eq. (2) obtained by fitting the experimental ARPES band dispersions
in (Pb,Sn)Te\cite{ARSuYang},
SnTe\cite{ARTanaka,ARTanaka2} and (Pb,Sn)Se\cite{ARDziawa,ARMadhabN}.
\{$m$, $m^\prime$, \} are given in units of eV, and
\{$v_{x\pm}$, $v_{2x}$,\ $v_{y\pm}$, $v_{2y}$\}
in units of eV$\cdot$\r{A}.}
\begin{tabular*}{0.45\textwidth}{@{\extracolsep{\fill}}| l | c | c | c | r|}
\hline
\cline{1-4}
\noalign{\smallskip}
  &   Pb$_{1-x}$Sn$_x$Te&   SnTe   &   \multicolumn{2}{c|}{Pb$_{1-x}$Sn$_{x}$Se}\\
  &   x=0.4   &      &   x=0.23 & x=0.3\\
\hline
$m$          &  -0.30  &  -0.30  &  -0.062  &  -0.056  \\
$m^\prime$   &  -0.15  &  -0.15  &  -0.036  &  -0.026  \\
$v_{x\pm}$     &  -2.3   &  -2.3   &  -2.55 &  -2.58    \\
$v_{2x}$     &   0.0   &   0.0   &  -0.64   &  -0.32  \\
$v_{y\pm}$     &    -5    &   -6.5   &   -3.55  &   -3.28   \\
$v_{2y}$     &    0    &    0    &     0      &     0 \\
\hline
\hline
\end{tabular*}
\label{table1}
\end{center}
\vspace{0.6cm}
\end{table}


\section{Discussion and conclusion}
In general, topological crystalline insulators should harbor
distinct classes with positive and negative mirror Chern numbers.
The schematic diagrams of possible spin textures around the lower 
Dirac cones are given in Fig.~4. In Fig.~4a, a pair of Dirac cones 
with counter-clockwise spin texture on the horizontal mirror line 
gives the mirror Chern number n$_{m}=-2$. The opposite case with a 
clockwise spin texture in Fig.~4b would give the mirror Chern number 
n$_{m}=2$. By comparing the spin-texture obtained from our 
first-principles calculations in Fig.~3, we can conclude that the 
mirror Chern number of SnTe is $-2$. For Bi$_{2}$Se$_{3}$ system 
shown in Fig.~4c, on a mirror line, only a single Dirac cone 
appears at the center of the hexagonal Brillouin zone with the mirror 
Chern number n$_{m}=-1$. In Fig.~4d, if the spin texture runs 
clockwisely, the system has the mirror Chern number n$_{m}=1$. 
Since there are two coaxial Dirac cones in SnTe, where the Sn-like and 
Te-like Dirac cones exhibit the same spin charality as Bi$_{2}$Se$_{3}$, 
it follows that the mirror Chern number of SnTe is twice that of Bi$_{2}$Se$_{3}$.

\begin{figure}
\includegraphics[width=5cm]{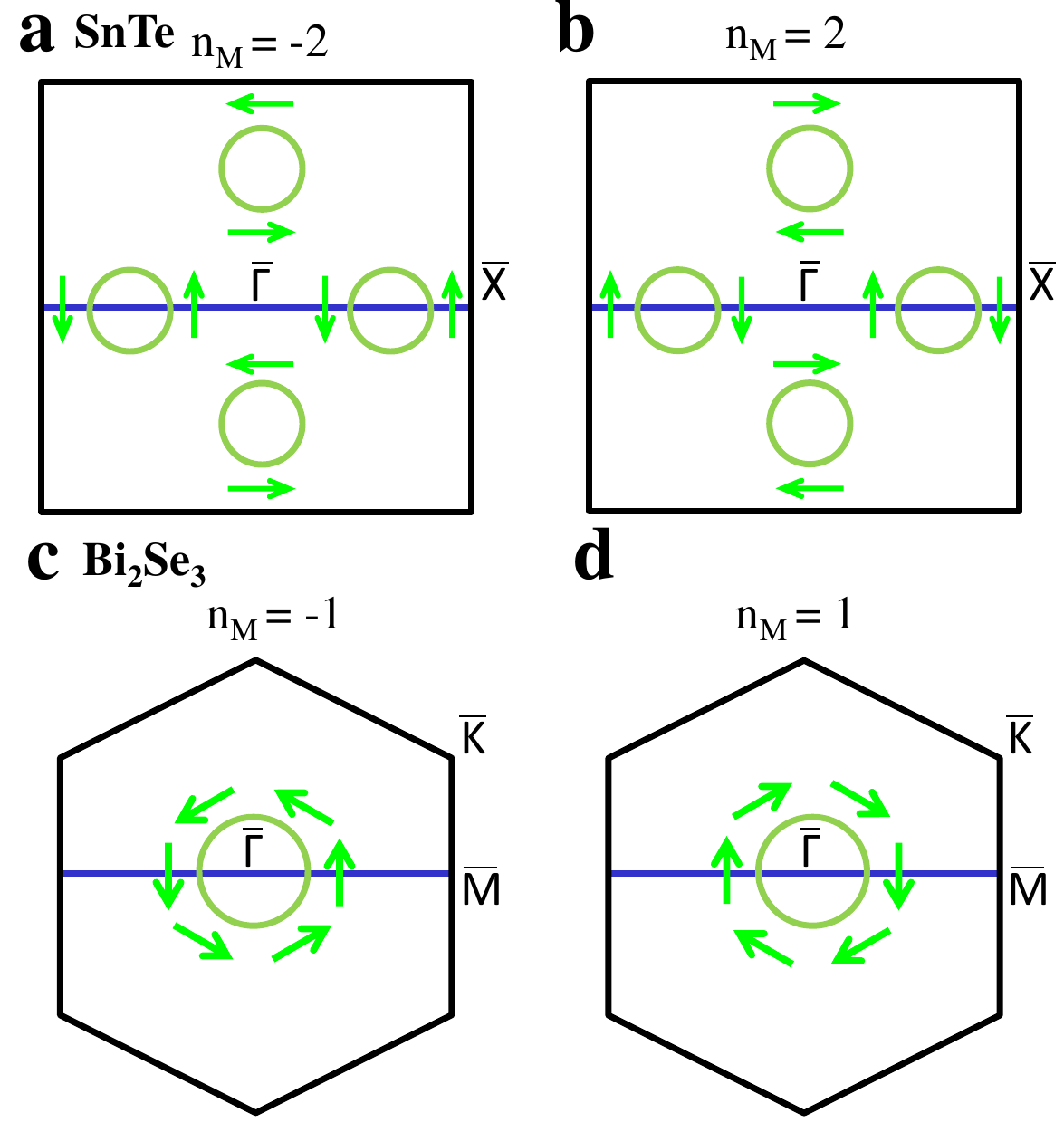}
\caption{(Color online)
{\bf Schematic diagrams for spin texture of the lower Dirac cones in distinct topological phases associated with various values of the mirror Chern number n$_m$.}
{\bf a.} n$_{m}=-2$; 
{\bf b.} n$_{m}=2$;
{\bf c.} n$_{m}=-1$;
{\bf d.} n$_{m}=1$.
Blue lines indicate the axis of mirror symmetry.
}
\label{f4}
\end{figure}

In conclusion, we have delineated the charge density distributions and spin textures of the [001] surface states on 
the topological crystalline insulator SnTe via first-principles calculations. We show that the SnTe surface states 
can be pictured as two interacting coaxial Dirac cones which intersect each other to form the Dirac points along 
the mirror symmetry line $\bar{\Gamma}$\={X}. From an examination of the charge distribution, we attribute the origin 
of the two coaxial Dirac cones to the two distinct atomic species, Sn and Te, in the material. The crystal and 
time-reversal symmetries guarantee that the out-of-the-plane spin polarization is zero.
The spin texture is dictated by the non-trivial Chern number n$_{m}=-2$ in SnTe, which is different from the known 
Z$_{2}$ topological insulators such as Bi$_{2}$Se$_{3}$.

\textbf{Acknowledgments:}
It is a pleasure to thank Liang Fu and Chen Fang for useful discussions.
The work at Northeastern and Princeton is supported by
the Division of Materials Science and Engineering, Basic
Energy Sciences, U.S. Department of Energy Grants DE-FG02-07ER46352,
DE-FG-02-05ER46200 and AC02-05CH11231, and benefited from theory support at the Advanced Light Source, Berkeley, and the
allocation of supercomputer time at NERSC and Northeastern University's
Advanced Scientific Computation Center (ASCC).
M.Z.H is supported by NSF-DMR-1006492 and DARPA-N66001-11-1-4110.
W.F.T is supported by the NSC in Taiwan under Grant No. 100-2112-M-110-001-MY2.


\begin{thebibliography}{99}

\bibitem{reviewHasan}
M. Z. Hasan, and C. L. Kane, 
{Rev. Mod. Phys.} {\bf 82}, 3045-3067 (2010).
%
\bibitem{reviewZhang}
X.-L. Qi, and S.-C. Zhang,  
{Rev. Mod. Phys.} {\bf 83}, 1057 (2011).
%
\bibitem{reviewMoore}
M. Z. Hasan, and J. E. Moore, 
{Annual Review of Condensed Matter Physics} {\bf 2}, 5578 (2011).
%

\bibitem{scHgTe}
M. Konig, H. Buhmann, L. W. Molenkamp, T. Hughes, C-X Liu, X-L Qi and S-C Zhang, J. Phys. Soc.
Japan {\bf 77}, 031007 (2008)

\bibitem{FuBi}
L. Fu and C.L. Kane, Phys. Rev. B {\bf 76}, 045302 (2007).

\bibitem{HsiehSb}
D. Hsieh, D. Qian, L. Wray, Y. Xia, Y. S. Hor, R. J. Cava and  M.Z. Hasan, Nature, {\bf 452}, 970 (2008).

\bibitem{HsiehBi}
Hsieh D \textit{et al.},  Nature 460, 1101 (2009)

\bibitem{FuTci}
L. Fu,
Phys. Rev. Lett. 106, 106802 (2011)

\bibitem{Natliang}
Timothy H. Hsieh \textit{et al.}, Nature Communications {\bf 3}, 982 (2012)

\bibitem{ARSuYang}
Su-Yang Xu \textit{et al.}, Nature Communications {\bf 3}, 1192 (2012) 

\bibitem{ARTanaka}
Y. Tanaka \textit{et al.}, Nature Physics {\bf 8}, 800 (2012)

\bibitem{ARDziawa}
P. Dziawa \textit{et al.}, Nature Materials {\bf 11}, 1023 (2012)

\bibitem{PRBTeo}
Teo, J. C. Y. , Fu, L., and Kane, C.L.  Phys. Rev. B {\bf 78}, 045426 (2007).

\bibitem{TakahashiPRL}
R. Takahashi and S. Murakami
Phys. Rev. Lett. {\bf 107}, 166805 (2011).

\bibitem{paw}
G. Kresse and D. Joubert, Phys. Rev. B {\bf 59}, 1758 (1999).

\bibitem{vasp}
J. P. Perdew, K. Burke and M. Ernzerhof, Phys. Rev. Lett. {\bf 77}, 3865 (1996)

\bibitem{gga}
G. Kresse and J. Hafner, Phys. Rev. B {\bf 48}, 13115 (1993)

\bibitem{lasnte}
R. F. Bis and J. R. Dixon, J. Appl. Phys. {\bf 40}, 1918 (1969)

\bibitem{foot1}
C. Fang {\it et al.}\cite{ARFang} considered a 
similar approach but without any spin information.

\bibitem{ARFang}
C. Fang \textit{et al.},
arXiv:1212.3285

\bibitem{foot2}
The picture of two degenerate Dirac cones given in the recent work 
by J. Liu {\it et al}.\cite{Fukpmodel} can be 
obtained via a unitary transformation of our Hamiltonian, 
and is thus equivalent to our $k\cdot p$ model.

\bibitem{Fukpmodel}
J. Liu, W. Duan and L. Fu, arXiv:1304.0430.

\bibitem{ARTanaka2}
Y. Tanaka \textit{et al.},
arXiv:1301.1177

\bibitem{ARMadhabN}
M. Neupane \textit{et al.},
in preparation

\end{thebibliography}
\end{document}